%
%
\magnification=\magstep1
\baselineskip=11pt plus .1pt minus .1pt
\hsize=12.5truecm
\vsize=19.0truecm  
\hfuzz=5pt\vfuzz=5pt
\tolerance=1000
\overfullrule=0pt
\parskip=0pt
\abovedisplayskip=3 mm plus6pt minus 4pt
\belowdisplayskip=3 mm plus6pt minus 4pt
\abovedisplayshortskip=0mm plus6pt minus 2pt
\belowdisplayshortskip=2 mm plus4pt minus 4pt
\predisplaypenalty=0
\clubpenalty=10000
\widowpenalty=10000
\parindent=2em
%
%
\font\pgnumfont=cmr9
\font\headlinefont=cmti9
 \font\titlefont=cmbx10
\font\authorfont=cmr10
\font\addressfont=cmti9
\font\datefont=cmr9
\font\sumfont=cmr9

\font\absfont=cmbx9
\font\secfont=cmr10
\font\subsecfont=cmti10
\font\subsubsecfont=cmr10
\font\figfont=cmr9
\font\figheadfont=cmbx9
\font\tabfont=cmr9
\font\tabheadfont=cmbx9
\font\mainfont=cmr10
\font\petitrm=cmr9

%
%
%
\newtoks\TITLE \newtoks\AUTHOR \newtoks\ADDRESS \newtoks\SUMMARY
\newdimen\sumindent \sumindent=\parindent
\newtoks\KEYWORDS \newtoks\SUBMITTED \newtoks\ACCEPTED
\newtoks\SENDOFF
%

%
%
\newtoks\firstpage
\let\firstpage=Y
\newtoks\AUTHORHEAD \newtoks\ARTHEAD \newtoks\VOLUME \newtoks\PAGES
\if!\the\AUTHORHEAD!\AUTHORHEAD={\the\AUTHOR}\fi
\if!\the\ARTHEAD!\ARTHEAD={\the\TITLE}\fi
\footline={\hfil}
\headline={\ifodd\pageno\rightheadline \else\leftheadline\fi}
\def\leftheadline{\if Y\firstpage\firsthead\global\let\firstpage=N
  \else\lefthead\fi}
\def\rightheadline{\if Y\firstpage\firsthead\global\let\firstpage=N
  \else\righthead\fi}
\def\lefthead{\pgnumfont\number\pageno\hfil\headlinefont\the\AUTHORHEAD}
\def\righthead{\headlinefont\the\ARTHEAD\hfil\pgnumfont\number\pageno}
\def\firsthead{\headlinefont Baltic Astronomy,~vol.\the\VOLUME,
\the\PAGES,~\the\year .\hfil}
\voffset=2\baselineskip 
%

\newdimen\oldbaselineskip \oldbaselineskip=\baselineskip
\def\test#1{\newlinechar=`@\if!\the#1! \message{#1 not given@}\fi}%
\def\printheader{
  \parindent=0pt
  \null\vskip1.cm
  \test{\TITLE}
  \vbox{\baselineskip=15pt
    \titlefont\the\TITLE
    }
  \vskip8mm plus8mm
  \test{\AUTHOR}
  \authorfont\the\AUTHOR
  \vskip2mm
  \test{\ADDRESS}
  \addressfont\the\ADDRESS
  \vskip2mm
  \test{\SUBMITTED}
  \line{\datefont Received \the\SUBMITTED
    \if!\the\ACCEPTED!\else, accepted \the\ACCEPTED\fi.\hfill}
  \vskip4mm plus4mm
  \vbox{\leftskip=\sumindent\parindent=0pt
    \parskip=5pt
    \absfont Abstract.
    \test{\SUMMARY}
    \sumfont\the\SUMMARY\par
    \absfont Key words:
    \test{\KEYWORDS}
    \sumfont\the\KEYWORDS\par
    }
  \sumfont
  \if!\the\SENDOFF!\else\footnote{}{Send offprint requests to:
 \the\SENDOFF}\fi
  \parindent=2em
  }
%
%
\newdimen\uppergap \newdimen\lowergap
\uppergap=5mm \lowergap=3mm
\newdimen\secind \newdimen\subsecind \newdimen\subsubsecind
\setbox0=\hbox{\secfont 9. }\secind=\wd0
\setbox0=\hbox{\subsecfont 9.9. }\subsecind=\wd0
\setbox0=\hbox{\subsubsecfont 9.9.9. }\subsubsecind=\wd0
\def\section#1{\goodbreak\par\vskip\uppergap
  \noindent\hangindent\secind\hangafter=1\secfont#1
  \vskip\lowergap\mainfont\par\nobreak}
\def\subsection#1{\goodbreak\par\vskip\uppergap
  \noindent\hangindent\subsecind\hangafter=1\subsecfont#1
  \vskip\lowergap\mainfont\par\nobreak}
\def\subsubsection#1{\goodbreak\par\vskip\uppergap
  \noindent\hangindent\subsubsecind\hangafter=1\subsubsecfont#1
  \vskip\lowergap\mainfont\par\nobreak}
%
%
\def\WFigure#1#2#3{\goodbreak\midinsert\vbox{
  \null\centerline{#2}\vskip1.5truemm
  \figheadfont\indent Fig.~#1.\figfont\ #3
  \par\mainfont
  }\endinsert}
%

%

%

%

%
\newdimen\tabind
\setbox0=\hbox{\tabheadfont Table 55.} \tabind=\wd0

%
%
\def\References{\vskip\uppergap
\line{\secfont REFERENCES\hfill}
  \vskip0.8\lowergap
 \petitrm
  }
\def\ref{\goodbreak
\hangindent12pt\hangafter=1
\noindent\ignorespaces}
\def\endref{\egroup}
%
%
\def\byebye{\egroup\par\vfill\supereject\end}
%
%

%
%

\def\utw{\smash{\rlap{\lower5pt\hbox{$\sim$}}}}
\def\udtw{\smash{\rlap{\lower6pt\hbox{$\approx$}}}}

\font\tbold=cmbx9
\font\tabfont=cmr9
\def\tablerule{\noalign{\vskip.9ex}\noalign{\hrule}\noalign{\vskip.7ex}}
\def\huad{\vrule height0pt depth0pt width5pt}  


\def\ddown{\lower2.5ex\hbox}
\def\ddow{\lower1.7ex\hbox}
\def\down{\lower1ex\hbox}
\def\uppp{\raise1ex\hbox}
\def\dnnn{\lower1ex\hbox}
\def\uuppp{\raise2ex\hbox}

\def\(o-c){$O-C$}


\def\angstr{A\kern-.56em\raise1.9ex\hbox{$\scriptscriptstyle\circ$}$\,$}

\newdimen\free\newdimen\shift
\def\Entry#1#2#3{\par\goodbreak\smallskip%
  \setbox1=\vbox{\advance\hsize by-10mm\parindent=0pt
    \def\\{\par}%
    \it#1. \rm#2}
  \line{\box1\hfill#3}\smallskip
}%
\newdimen\savesize

\def\shiftfigure #1#2#3#4#5{
    \vbox to #2 { \ifodd #5 \rightskip#4 \else\leftskip#4 \fi
                  \null\vfil
                  \figheadfont Fig.~#1.\figfont #3
                  \medskip
                }
                          }

\year1999

\input psfig.sty
\TITLE={THE ASTEROSEISMOLOGY METACOMPUTER}
\VOLUME={8}
\PAGES={XXX--XXX}              
\pageno=1                      
\AUTHORHEAD={T.S.~Metcalfe and R.E.~Nather}
\AUTHOR={T.S.~Metcalfe and R.E.~Nather}
\ARTHEAD={The Asteroseismology Metacomputer}

\ADDRESS={
\item{ }Department of Astronomy, University of Texas, Austin, TX 78701 U.S.A.}

\SUMMARY={
We have developed a specialized computational instrument for fitting
models of pulsating white dwarfs to observations made with the Whole Earth
Telescope. This metacomputer makes use of inexpensive PC hardware and free
software, including a parallel genetic algorithm which performs a global
search for the best-fit set of parameters.
}

\KEYWORDS={instrumentation: miscellaneous -- methods: numerical -- stars: white dwarfs}
\SUBMITTED={September 1, 1999}
\printheader

\section{1. INTRODUCTION}
       
White dwarf asteroseismology offers the opportunity to probe the structure
and composition of stellar objects governed by relatively simple
principles. The observational requirements of asteroseismology have been
addressed by the development of the Whole Earth Telescope (WET), but the
analytical procedures need to be refined to take full advantage of the
possibilities afforded by the WET data.

The adjustable parameters in our computer models of white dwarfs presently
include the total mass, the temperature, the H and He layer masses, the core 
composition, and the transition zone thicknesses. Finding a proper set of 
these to provide a close fit to the observed data is difficult. The current
procedure is a cut-and-try process guided by intuition and experience, and
is far more subjective than we would like. Objective procedures for
determining the best-fit model are essential if asteroseismology is to
become a widely-accepted and reliable astronomical technique.  We must be
able to demonstrate that, within the range of different values the model
parameters can assume, we have found the only solution, or the best one if
more than one is possible. To address this problem, we are applying a
search-and-fit technique employing a genetic algorithm (GA), which can
explore the myriad parameter combinations possible and select for us the
best one, or ones (cf.~Goldberg 1989, Charbonneau 1995, Metcalfe \& Nather
1999).

Although genetic algorithms are more efficient than other comparably
global techniques, they are still quite demanding computationally. To be
practical, the GA-based fitting technique requires a dedicated instrument
to perform the calculations. Over the past year, we have designed and 
configured such an instrument---an isolated network of 64 minimal PCs 
running Linux. Since the structure of a GA is very conducive to 
parallelization, this metacomputer allows us to run our code much faster 
than would otherwise be possible.

\section{2. HARDWARE}

In January 1998, around the time that the idea of commodity parallel
processing started getting a lot of attention, we were independently
designing a metacomputer of our own. Our budget was modest, so we set out
to get the best performance possible per dollar without restricting the
ability of the machine to solve our specific problem.

The original Beowulf cluster (Becker et al.~1995), which we didn't know
about at the time, had a number of features which, though they contributed
to the utility of the machine as a multi-purpose computational tool, were
unnecessary for our particular problem. We wanted to use each node of the
metacomputer to run identical tasks with small, independent sets of data.
The results of the calculations performed by the nodes consisted of just a
few numbers which only needed to be communicated to the master process,
never to another node. Essentially, network bandwidth was not an issue
because the computation to communication ratio of our application was
extremely high, and hard disks were not needed on the nodes because our
problem did not require any significant amount of data swapping. In the
end we settled on a design including one master server augmented by
minimal nodes connected by a simple 10base-2 network (see Figure 1).

\WFigure{1}{\psfig{figure=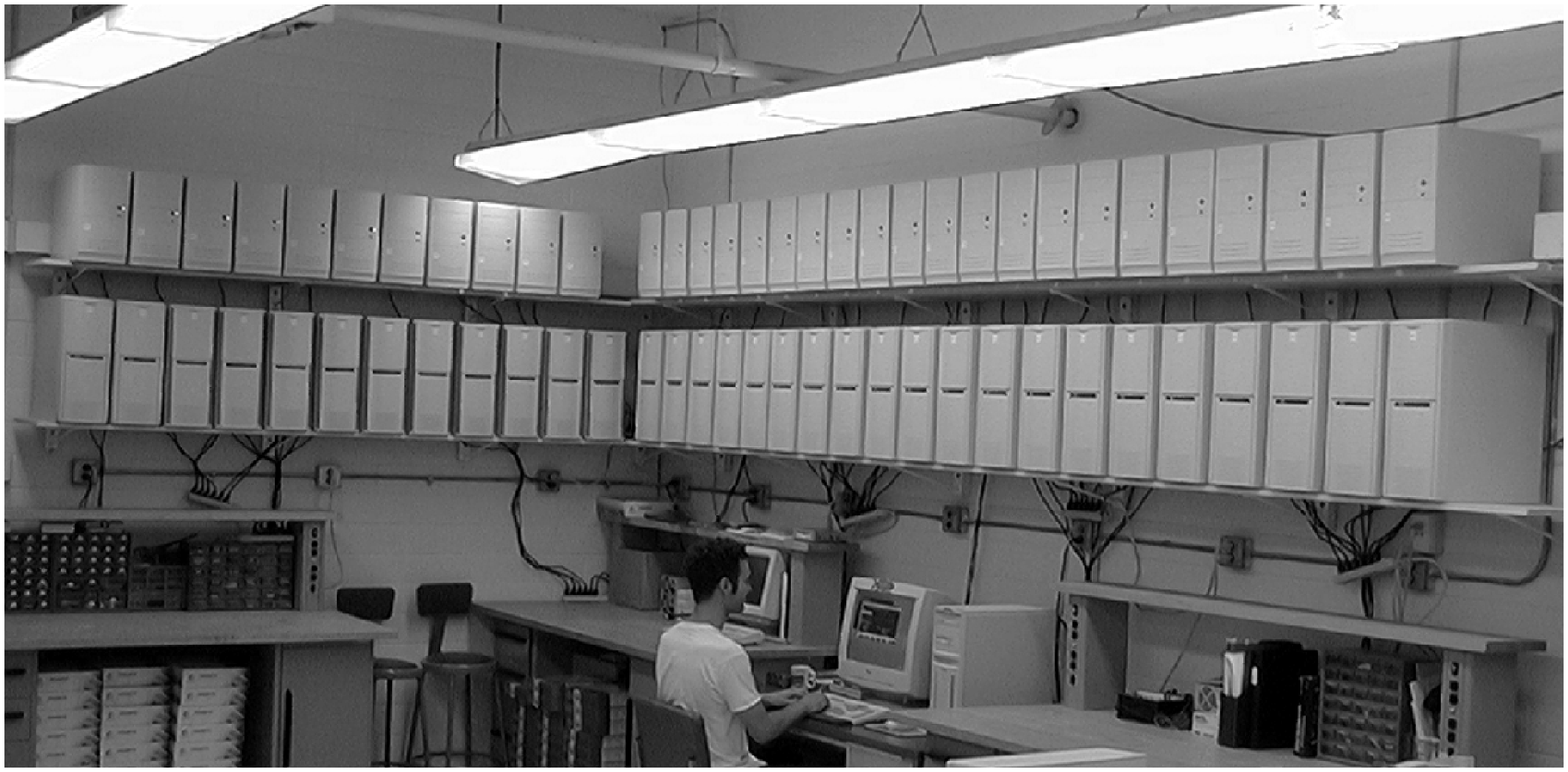,width=12truecm,angle=0,clip=}}
{The 64 minimal nodes of the metacomputer on shelves surrounding the 
master computer.}

The master computer is a Pentium-II 333 MHz system with three NE-2000
compatible network cards, each of which drives 1/3 of the nodes on a
subnet. Since a single ethernet card can handle up to 30 devices, no
repeater was necessary.

The slave nodes were assembled from components obtained at a discount
computer outlet. Each node consists of an ATX tower case with a
motherboard, processor and fan, a single 32 MB SDRAM, and an NE-2000
compatible network card with a custom made boot-EPROM. The nodes are
connected in series with 3-ft ethernet coaxial cables. Half of the nodes
contain Pentium-II 300 MHz processors, while the other half are AMD K6-II
450 MHz chips. The total cost of the system was around \$25k, but it could
be built for considerably less today, and less still tomorrow.

\section{3. SOFTWARE}

To make the metacomputer work, we relied on the open-source Linux
operating system and software tools. We programmed the EPROMs with Gero
Kuhlmann's NETBOOT package to allow each node to download and mount
an independent Linux filesystem on a small ramdisk partition. We used Tom
Fawcett's YARD package to create the filesystem, and we included in
it a pared down version of the PVM software developed at Oak Ridge
National Laboratory (Geist et al.~1994).

Finally, we incorporated the message passing routines of the PVM
library into PIKAIA, a general purpose public-domain GA developed by
Charbonneau (1995), and we modified the white dwarf evolution and
pulsation codes (see Wood 1990, Bradley 1993, Montgomery 1998) to allow 
reliable and automated calculation of the normal modes of oscillation for 
white dwarf stars with a wide range of masses, temperatures, and other
parameters.

\section{4. BENCHMARKS}

Measuring the absolute performance of the metacomputer is difficult 
because the result strongly depends on the fraction of Floating-point Division
operations (FDIVs) used in the benchmark code. Table 1 lists four different 
measures of the absolute speed in Millions of FLoating-point Operations Per 
Second (MFLOPS). 

$$\vbox{\tabfont
\halign{
\hfil\huad #\hfil\huad&           
\hfil\huad #\hfil\huad&           
\hfil\huad #\hfil\huad&           
\hfil\huad #\huad\hfil\cr         
{\tbold Table 1.}
&\multispan{3}{\ The absolute speed of the metacomputer.\hfil}\cr
\tablerule
Benchmark & P-II 300 MHz & K6-II 450 MHz & Total Speed \cr
\tablerule
MFLOPS(1) & 80.6         & ~65.1         & 4662.4      \cr
MFLOPS(2) & 47.9         & ~67.7         & 3699.2      \cr
MFLOPS(3) & 56.8         & 106.9         & 7056.0      \cr
MFLOPS(4) & 65.5         & 158.9         & 7180.8      \cr
\tablerule
}}
$$

The code for MFLOPS(1) is essentially scalar---that is, vector processor 
performance will reflect scalar performance which will lie far below 
expected vector performance. Also, the percentage of FDIVs (9.6\%) is 
considered somewhat high. The code for MFLOPS(2) is fully vectorizable. The 
percentage of FDIVs (9.2\%) is still somewhat on the high side. The code 
for MFLOPS(3) is also fully vectorizable. The percentage of FDIVs (3.4\%) 
is considered moderate. The code for MFLOPS(4) is fully vectorizable, but 
the percentage of FDIVs is zero.

We feel that MFLOPS(3) provides the best measure of the expected performance
for the white dwarf code, because of the moderate percentage of FDIVs.
Adopting this value, we have achieved a price to performance ratio near 
\$3.50/MFLOPS.
\vskip3mm

ACKNOWLEDGEMENTS.~We would like to thank Gary Hansen for donating the 32 
K6-II 450 MHz processors through an arrangement with AMD. This work was
made possible by a grant from the National Science Foundation.

\vfill\eject

\References
\ref 
Becker, D., Sterling, T., Savarese, D., Dorband, J., Ranawak, U., and
Packer, C. 1995, Beowulf: A Parallel Workstation for Scientific Computation, in Proceedings of the International Conference on Parallel Processing (New York: Institute of Electrical and Electronics Engineers).
\ref 
Bradley, P. 1993, Ph.D. Thesis, University of Texas at Austin.
\ref 
Charbonneau, P. 1995, ApJS, 101, 309. 
\ref 
Geist, A., Beguelin, A., Dongarra, J., Jiang, W., Manchek, R., and Sunderam, V. 1994, PVM: Parallel Virtual Machine, A Users' Guide and Tutorial for Networked Parallel Computing, (Cambridge: MIT Press).
\ref 
Goldberg, D. 1989, Genetic Algorithms in Search, Optimization, and Machine Learning, (Reading, MA: Addison Wesley).
\ref 
Metcalfe, T.S. and Nather, R.E. 1999, Linux Journal, 65, 58.
\ref 
Montgomery, M. 1998, Ph.D. Thesis, University of Texas at Austin.
\ref 
Wood, M. 1990, Ph.D. Thesis, University of Texas at Austin.

\bye